\documentclass[11pt,a4paper]{iopart}
\usepackage[utf8]{inputenc}
\expandafter\let\csname equation*\endcsname\relax

\expandafter\let\csname endequation*\endcsname\relax 
\usepackage{amsmath}
\usepackage{amsfonts}
\usepackage{amssymb}
\usepackage{bm}
\usepackage{graphicx}
\usepackage{iopams}
\usepackage{todonotes}
\usepackage{hyperref}
\usepackage[normalem]{ulem} 
\usepackage{soul}

\def\bea{\begin{eqnarray}}
\def\eea{\end{eqnarray}}

\def\la{\langle}
\def\ra{\rangle}

\newcommand{\rt}{\textrm{RT}}

\begin{document}

\title{Long time behavior of run-and-tumble particles in two dimensions}

\author{Ion Santra$^{1}$, Urna Basu$^{2}$, Sanjib Sabhapandit$^{1}$}

\address{$^{1}$Raman Research Institute, Bengaluru 560080, India}

\address{$^{2}$S. N. Bose National Centre for Basic Sciences, Kolkata 700106, India}

\begin{abstract}
We study the long-time asymptotic behavior of the position distribution of a run-and-tumble particle (RTP) in two dimensions in the presence of translational diffusion and show that the distribution at a time $t$ can be expressed as a perturbative series in $(\gamma t)^{-1}$, where $\gamma^{-1}$ is the persistence time of the RTP. We show that the higher order corrections to the leading order Gaussian distribution generically satisfy an inhomogeneous diffusion equation where the source term depends on the previous order solutions.  The explicit solution of the inhomogeneous equation requires the position moments, and we develop a recursive formalism to compute the same. We find that the subleading corrections undergo shape transitions as the translational diffusion is increased.
\end{abstract}

\section{Introduction}
Active particles form a class of nonequilibrium systems that can self-propel by consuming energy from their surroundings~\cite{BechingerRev,SriramRev,Romanczuk}. They are found abundantly in nature, ranging from birds in a flock~\cite{Giardina2014,Bialek2012}, fish schools~\cite{Partridge1982,Guttal2020}, micro-organisms like bacteria~\cite{bergbook} to artificial objects like Janus particles and micro and nano-robots used for targeted drug delivery~\cite{Sano2010,Bechinger2012}. An important theoretical approach 
involves minimal stochastic modeling of active particles, mimicking the different types of self-propelled dynamics seen in nature~\cite{Franosch2016,Mori2021,Rosalba2021,Zhang2022,Santra2020_reset,Shee2020,Squarcini2022,
Hartmann2020,Singh2020,Maes2018,Banerjee2020}. These models typically describe the motion of an overdamped particle with a propulsion velocity that is correlated in time---effectively generating a persistent motion. The propulsion velocity has a stochastic dynamics of its own, which differ based on the kind of active motion it is used to model~\cite{Berg1972,Tailleur2008,solon2015active,Malakar2017,Santra2020,Leonardo2014,Martin2021,
Howse2007,Basu2018,Thutupalli2019,Santra2021,angelani2022orthogonal}. Active motions break the detailed balance condition, violate the fluctuation-dissipation relations, exhibit non-diffusive scalings~\cite{Sevilla2014,Basu2018,Santra2021} and non-Boltzmann stationary states in presence of confining potentials~\cite{Pototsky2012,Basu2019,Dhar2019,Malakar2020,Basu2020,Santra2021_trap}, as well as unusual first-passage properties~\cite{Mori2020_prl,Woillez2019,Basu2018,Santra2021}, etc. Due to the inherent nonequilibrium nature of the dynamics, the exact analytical treatment is non-trivial. In fact, there is no general formalism to understand these active dynamics, and one has to approach the different models of self-propulsion in different ways to extract their statistical properties. Recently, however, it was shown~\cite{Santra_2022} that the long-time dynamics of active particles show certain universal behavior irrespective of the specific dynamics of the stochastic propulsion velocity, namely, the leading order position distribution is a Gaussian with a diffusive scaling and the subleading corrections to this leading order Gaussian follow an inhomogeneous diffusion equation. The specificity of the different active dynamics enters through the source term only, which, at each order, depends on the previous order solutions.

One of the earliest and most popular models of active motion is the run-and-tumble particles (RTP)~\cite{Berg2004coli} that mimics the dynamics of bacteria like {\it E. coli}. The motion consists of segment of `run' phases, where the particle propels itself along an internal orientation at a constant speed, intermittently interrupted by `tumbling' events where the internal orientation changes randomly. The simplest and the most studied version is the one-dimensional RTP, where the internal orientation can flip between two possible values $\pm 1$. The equations of motion describing this dynamics correspond to the Telegraphers equations~\cite{masoliver1992solutions,masoliver1993solution} and are very well-studied in the literature. However, the usual observations of run-and-tumble processes in nature are in higher dimensions. In fact, in the image tracking experiments, one usually looks at the projected motion of RTP in two-spatial dimensions~\cite{Berg1972}. In two dimensions, the internal orientation is characterized by a continuous angle $\theta$, which evolves through a jump process~\cite{berg1975bacterial}.
 The exact position distribution of this dynamics has been previously calculated in~\cite{Santra2020,martens2012probability}---where it was found that, at early times, starting from a randomized initial orientation, the position distribution at early times is concentrated along a ring around the origin that grows in time ballistically. In contrast, at large times, the typical [$x\sim O(\sqrt{t})$] fluctuations are Gaussian, and the signatures of activity are encoded in the atypical fluctuations $x\sim O(t)$ that are characterized by a large deviation function~\cite{Santra2020}. 
 
 To understand the effects of the activity of an RTP at late times via the large deviation functions experimentally is challenging as the $O(t)$ events are rare.
In this paper, we study the $O(\sqrt{t})$ fluctuations beyond Gaussian, of the two-dimensional RTP at long-times using the formalism developed in~\cite{Santra_2022}.
Starting from the Fokker-Planck equation, we show that at the leading order, the position distribution satisfies a diffusion equation, which yields a leading order Gaussian position distribution. Further, we show that the subleading contributions follow inhomogeneous diffusion equations at each order. We solve the first few of these explicitly to obtain the corrections to the leading order Gaussian distribution in the typical region.  
 These corrections, of $O(\sqrt{t})$, are more accessible experimentally and are thus better markers of the signature of activity at large times from an experimental point of view.

The simplest model outlined above does not take into account the translational diffusion due to the thermal fluctuations of the medium, which can be important in realistic situations. Here we also study the long-time behavior of two-dimensional RTP in the presence of translational diffusion, for which the exact position distribution is not known. We calculate the subleading corrections in this case and show that the universal structure of the long-time position distribution remains the same. Interestingly, however, the corrections undergo some interesting shape transitions at the origin as the translational diffusion is enhanced.

The paper is organized as follows. We first describe the two-dimensional RTP dynamics and the associated Fokker-Planck equation without the translational diffusion in Sec.~\ref{secmodel}. The position moments are calculated in Sec.~\ref{secmoments}. The leading order position distribution at long-times and its subleading corrections are computed in Sec.~\ref{secpdf}. We incorporate the translational diffusion in Sec.~\ref{sec:thermal} and study the position distribution at long-times by calculating the leading order distribution and its subleading corrections. Finally, we conclude with some general remarks in Sec.~\ref{conclusion}.

\section{The model and the Fokker-Planck equation}\label{secmodel}
The run-and-tumble dynamics describes the overdamped motion of a particle that `runs' at a constant speed $v_0$ along an internal orientation that changes stochastically, via `tumbling' at a constant rate $\gamma$. In two dimensions, the orientation is characterized by a unit vector $\bm{\hat{n}}=(\cos\theta,\sin\theta)$ and the tumbling results in $\theta\to\theta'$, where $\theta'$ is chosen uniformly from $[0,2\pi]$. The dynamics of this two dimensional RTP is described by the following overdamped Langevin equation,
\begin{align}
\dot{\vec x}=\vec{v}(t), \quad\text{with  }v_1(t)=v_0\cos\theta(t),~~v_2(t)=v_0\sin\theta(t).\label{e:langevin}
\end{align}
In reality, there can also be a Brownian noise in addition to the active noise $v(t)$. However, in many practical situations, for example, a bacterium swimming in water at room temperature, the effect of this thermal noise is negligible---in one second, a living E. coli moves about $20-30\,\mu$m, whereas the typical displacement of a dead one due to thermal noise is about $1\,\mu$m. Therefore, we first consider the scenario described by \eref{e:langevin}, ignoring the effect of thermal noise. We study the system with the thermal noise later in Sec.~\ref{sec:thermal}.

We consider the initial condition where the particle starts at the origin with an orientation chosen uniformly from $[0,2\pi]$, implying that $\la\cos\theta(t)\ra=\la\sin\theta(t)\ra=0$. Consequently, the components of stochastic velocity have zero mean. Furthermore, if  there is at least one tumbling event during the interval $[t,t']$, then $\theta(t)$ and $\theta(t')$ are independent. On the other hand, the orientation $\theta(t)=\theta(t')$ remains unchanged if there are no tumbling events during $[t,t']$. Consequently,
\begin{equation}
\la\cos\theta(t)\cos\theta(t')\ra_c=\la\sin\theta(t)\sin\theta(t')\ra_c=\begin{cases}0\quad &\text{for at least one tumbling event in }[t,t'],
\\\displaystyle\frac{1}{2}\quad &\text{for no tumbling events in }[t,t'],
\end{cases}
\end{equation}
where the subscript $c$ indicates conditional expectations.
The probability that there is no tumbling event within the duration $|t-t'|$ is $\exp\left(-\gamma|t-t'|\right)$. Therefore, the components of the stochastic velocity have exponentially decaying autocorrelations,
\begin{align}
\la v_x(t)v_x(t') \ra=\la v_y(t)v_y(t') \ra=\frac{v_0^2}{2}e^{-\gamma |t-t'|}.\label{vautoc}
\end{align}
It is evident from the above equation that the stochastic velocity becomes weakly correlated at times $|t-t'|\gg\gamma^{-1}$, the persistence time, and hence, by appealing to the central limit theorem, we expect a Gaussian distribution for the typical fluctuations with $\la x_1^2(t)=\la x_2^2(t)\ra\ra\simeq 2D_\rt t$ where $D_\rt=v_0^2/2\gamma$. However, corrections to the Gaussian distribution cannot be obtained from this heuristic argument.
In the following, starting from the Fokker-Planck equation, we rigorously derive this late-time diffusive behavior as well as the subleading corrections to it systematically.

The position distribution remains isotropic at all times when the initial orientation is chosen uniformly in $[0,2\pi]$. Therefore, it suffices to consider only $P_r(r,t)$, the distribution of the radial part $r=\sqrt{x_1^2+x_2^2}$, which is, in turn, related to the Cartesian marginal distribution $P(x_1,t)$ by $P_r(r,t)=2P(x_1=r,t)$. The Fokker-Planck equation for the joint distribution $P(x_1,\theta,t)$ is given by,
\begin{align}
\left[\frac{\partial }{\partial t}+v_0\cos\theta \frac{\partial}{\partial x_1}\right]P(x_1,\theta,t)=\gamma{\cal L}_\theta P(x_1,\theta,t) ,\label{pxthetat}
\end{align}
where ${\cal L}_\theta$ is the Markov operator corresponding to the $\theta$ dynamics,
\begin{align}
{\cal L}_\theta P(x_1,\theta,t)=-P(x_1,\theta,t)+\int_0^{2\pi}\frac{d\theta'}{2\pi}P(x_1,\theta',t).\label{ltheta}
\end{align}
To obtain the solution of \eref{pxthetat} in the long time regime $(t\gg \gamma^{-1}),$ it is convenient to introduce the scaled variable $y=x_1/\sqrt{D_{\rt}}=\sqrt{2\gamma}\,x_1/v_0$, such that the corresponding distribution $Q(y,\theta,t)$ satisfies,
\begin{align}
\left[\varepsilon^2\frac{\partial }{\partial t}+\varepsilon\sqrt{2}\cos\theta \frac{\partial }{\partial y}\right]Q(y,\theta,t)={\cal L}_\theta Q(y,\theta,t),
\label{fp-scaled}
\end{align}
where $\varepsilon^2=\gamma^{-1}$. In the following, we solve the above equation perturbatively, by treating $\varepsilon^2/t$ as a small parameter following the framework developed recently~\cite{Santra_2022}. To explicitly obtain the coefficient of the $(\varepsilon^2/t)^k$ term in the perturbative series, the knowledge of the position moments $\la y^{2k}(t)\ra$ is required. Hence, we first develop a recursive procedure to compute the position moments in the next section.

\section{Moments}\label{secmoments}
To compute the position moments of the two-dimensional RTP, it is convenient to start with the correlation functions,
\begin{align}
M(k,n,t)=\int_{-\infty}^{\infty}dy\int_0^{2\pi}d\theta\,y^k\,\cos(n\theta)\, Q(y,\theta,t),\label{mkn_def}
\end{align}
where $k\geq0,\,n\geq0$ are integers. Note that, $M(k,0,t)=\la y^k(t)\ra $, i.e.,  the position moments are obtained by putting $n=0$ in \eref{mkn_def}. The normalization condition of the joint distribution $Q(y,\theta,t)$ and the fact that $\la \cos(n\theta)\ra=0$ for $n>0$, leads to the condition,
\begin{align}
M(0,n,t)=\delta_{n,0}.\label{m0n}
\end{align}

The time evolution equations for $M(k,n,t)$ can be derived by multiplying both sides of the Fokker-Planck equation \eref{fp-scaled} by $y^k\,\cos(n\theta)$ and integrating over $y$ and $\theta$, which leads to, for $k>0$,
\begin{align}
\left(\varepsilon^2\frac{d}{dt}+1\right)M(k,n,t)&=\frac{\varepsilon k}{\sqrt{2}}\left[M(k-1,n-1,t)+M(k-1,n+1,t)\right]\text{ for }n>0,\label{mknt_d}\\
\text{ and }~\varepsilon\frac{d}{dt}M(k,0,t)&=k\sqrt{2}\,M(k-1,1,t)\label{mk0t_d}.
\end{align}
 Since we assume that the particle starts from origin at $t=0$, the recursive first-order differential equations \eref{mknt_d} and \eref{mk0t_d} must satisfy the initial condition $M(k,n,0)=0$ for $n>0$ and arbitrary $k$. Thus, formally, we have, from  \eref{mknt_d} and \eref{mk0t_d},
\begin{align}
M(k,n,t)=\frac{k}{\sqrt{2}\,\varepsilon}\int_0^t ds\, e^{-(t-s)/\varepsilon^2}\left[M(k-1,n-1,s)+M(k-1,n+1,s) \right] \text{ for }n>0,
\label{mkn}
\end{align}
 and
\begin{align}
M(k,0,t)=\frac{k\sqrt{2}}{\varepsilon}\int_0^t ds\, M(k-1,1,s),
\label{mk0}
\end{align}
for $n=0$, respectively. 
The correlation functions $M(k,n,t)$ can be computed recursively from \eref{mkn}, \eref{mk0} using the boundary condition \eref{m0n}.

\begin{figure}
\centering\includegraphics[width=0.5\hsize]{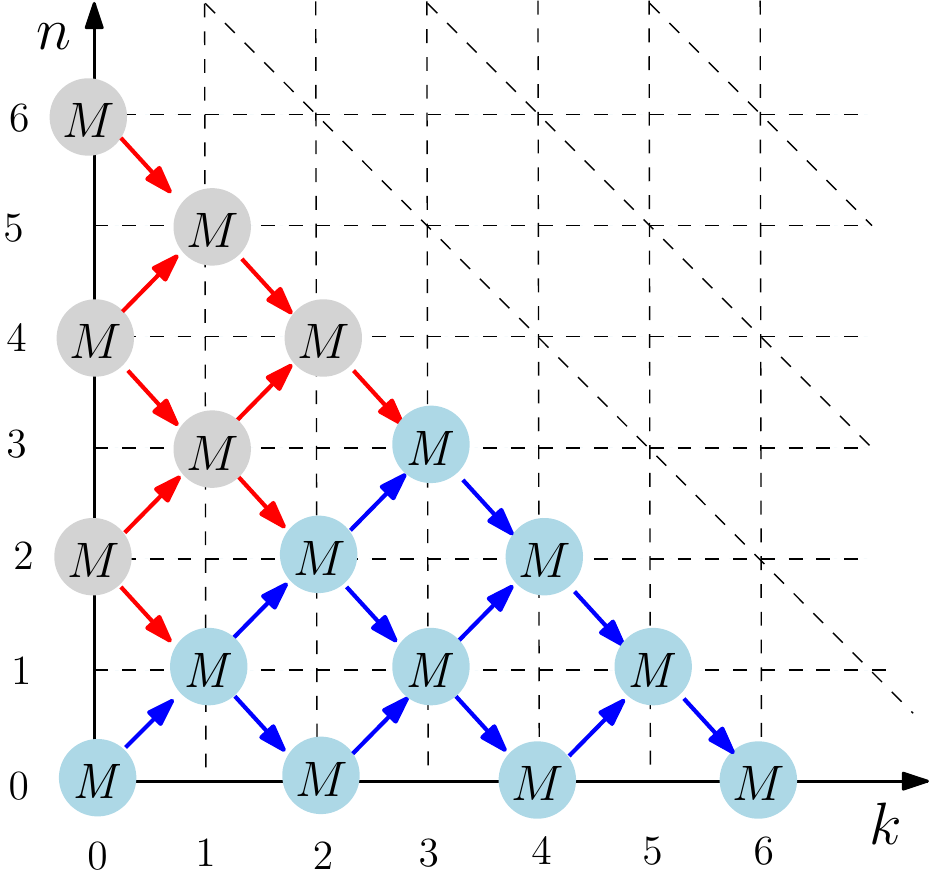}
\caption{Illustration of the recursive connection among the correlation functions $M(k,n,t)$ on the $(k,n)$ lattice. The blue discs indicate the points where the correlation function is non-zero, while the grey ones indicate the positions where $M(k,n,t)=0$.}\label{fig:mkn}
\end{figure}

Since the particle starts from the origin with $\theta$ chosen uniformly between $[0,2\pi]$, the odd position moments are always zero, i.e., $M(2k+1,0,t)=0$.  Moreover, equation \eref{mkn} implies that the set of correlation functions $M(k,n,t)$ form two independent networks, sitting on even and odd values of $k+n$ respectively. Hence, the condition \eref{m0n} along with the fact that $M(2k+1,0,t)=0$, implies that $M(k,n,t)=0$ for all odd $k+n$. Thus, to determine the non-zero position moments $M(2k,0,t)$, we need to consider the even $k+n$ network only [see \fref{fig:mkn} for a schematic representation]. It is further clear from \fref{fig:mkn} that the correlation functions $M(k,n,t)$ vanish for $n>k$. Consequently, the correlations on the line $n=k$ simplify to,
\begin{align}
M(k,k,t)=\frac{k}{\sqrt{2}\,\varepsilon}\int_0^t ds\, e^{-(t-s)/\varepsilon^2}\,M(k-1,k-1,s).\label{mkkt:1}
\end{align}
This integral recursion relation can be solved exactly to yield [see \ref{sec:mkkt} for details],
\begin{align}
M(k,k,t)=\left(\frac{\varepsilon}{\sqrt{2}}\right)^k k!\left[1-e^{-t/\varepsilon^2}\sum_{\nu=0}^{k-1}\frac{(t/\varepsilon^2)^\nu}{\nu!}\right].\label{mkkt}
\end{align}
For example, the first few diagonal terms are given by,
\begin{align}
M(1,1,t)&=\frac{\varepsilon}{\sqrt{2}}\left(1-e^{-t/\varepsilon^2}\right),\label{m11}\\
M(2,2,t)&=\varepsilon ^2-\left(t+\varepsilon ^2\right)e^{-t/\varepsilon ^2}, \\
M(3,3,t)&=\frac{3\varepsilon ^3}{\sqrt{2}}-\frac{3}{2\sqrt{2}\varepsilon} e^{-\frac{t}{\varepsilon ^2}} \left(t^2+2 t \varepsilon ^2+2 \varepsilon ^4\right).
\end{align}

Next, we calculate the first few moments explicitly starting from $\la y^2(t)\ra=M(2,0,t)$. Substituting $k=2$ in \eref{mk0}, we get,
\begin{align}
M(2,0,t)=\frac{2\sqrt{2}}{\varepsilon}\int_0^t ds\, M(1,1,s),\label{m20defn}
\end{align}
which, in turn, can be evaluated using \eref{m11}, to get,
\begin{align}
M(2,0,t)=2t-2\varepsilon^2(1-e^{-t/\varepsilon^2}).\label{m20t_f}
\end{align}
We can proceed in a similar manner to calculate the higher order position moments by substituting $k=2,\,4,\dotsc$ in \eref{mk0} and thereafter evaluating the terms appearing on the right hand side using \eref{mkn}. We evaluate the next two non-zero moments,
\begin{align}
M(4,0,t)=6t^2(2-e^{-t/\varepsilon^2})-36 \varepsilon^2 t+36\varepsilon^4(1-e^{-t/\varepsilon^2}),
\end{align} 
and
\begin{align}
M(6,0,t)=\frac{15t^4}{\varepsilon^2} e^{-t/\varepsilon^2}+120t^3-180\varepsilon^2 t^2(4-e^{-t/\varepsilon^2})+1800\varepsilon^4 t-1800\varepsilon^6(1-e^{-t/\varepsilon^2}).
\end{align} 
Even higher order position moments can be obtained systematically following the same procedure.

\section{Position distribution}\label{secpdf}
In this section, we derive the position distribution of the RTP at long times perturbatively. Integrating \eref{fp-scaled} over $y$ gives the Fokker-Planck equation for the marginal distribution of $\theta$, 
\begin{align}
\varepsilon^2\frac{\partial R(\theta,t)}{\partial t}={\cal L}_\theta R(\theta,t),\qquad\text{where~~}R(\theta,t)=\int_{-\infty}^{\infty}dy\, Q(y,\theta,t).
\end{align}
The operator ${\cal L}_\theta$, defined by \eref{ltheta}, has one non-degenerate eigenvalue $0$ with the corresponding eigenfunction $p_0=1/(2\pi)$. The remaining eigenfunctions $p_n(\theta)=\cos(n\theta)/\pi$ share a common eigenvalue $-1$.
The eigenfunctions obey the orthonormality relations,
\begin{align}
\int_0^{2\pi} d\theta\cos(m\theta)p_n(\theta)&=\delta_{m,n}~~ \text{and~~}
\int_0^{2\pi} d\theta\cos(\theta)\cos(m\theta)p_n(\theta)=\frac12(\delta_{n-1,m}+\delta_{n+1,m}).
\end{align}
Therefore,  $R(\theta,t)$, for a given initial condition $\theta(t=0)=\theta_0$, can be written as,
\begin{align}
R(\theta,t|\theta_0)=\frac{1}{2\pi}+e^{-t/\varepsilon^2}\sum_{n=1}^{\infty}\cos(n\theta_0) p_n(\theta),
\end{align}
where we have used the initial condition $R(\theta,0|\theta_0)=\delta(\theta-\theta_0)$ and the orthonormality condition. In fact, setting $t=0$ in the above equation  yields
 $\sum_{n=1}^{\infty}\cos(n\theta_0) p_n(\theta)=\delta(\theta-\theta_0) - (2\pi)^{-1}$, which leads to a simpler expression,
\begin{align}
R(\theta,t|\theta_0)=\frac{1}{2\pi}(1-e^{-t/\varepsilon^2}) + e^{-t/\varepsilon^2}\delta(\theta-\theta_0).
\end{align}
The second term on the right hand side comes from trajectories where $\theta$ has not tumbled, while the first term denotes the contributions from trajectories that have undergone at least one tumble event.
Evidently, the $\theta$ distribution reaches the stationary state $R(\theta,t\to\infty)=1/(2\pi)$. Moreover, if the initial orientation $\theta_0$ is chosen from the stationary state itself, then it remains stationary at all times, i.e., $\int_0^{2\pi} R(\theta,t|\theta_0) d\theta_0/(2\pi) = 1/(2\pi) $ ---which is the case considered here.

Since $\{p_n(\theta)\}$ form a complete basis, the joint distribution $Q(y,\theta,t)$ can be expanded as,
\begin{align}
Q(y,\theta,t)=\sum_{n=0}^{\infty} F_n(y,t)\,p_n(\theta),\label{qythetat}
\end{align}
where the series coefficients $F_n(y,t)=\int_{-\infty}^{\infty}d\theta\, Q(y,\theta,t)\cos(n\theta)$. Note that, since the $\theta$ distribution is stationary at all times, $\int_{-\infty}^{\infty}dy F_n(y,t)=\delta_{n,0}/(2\pi).$ 
Our goal is to find the marginal position distribution 
\begin{align}
\rho(y,t)=\int_0^{2\pi}d\theta\, Q(y,\theta,t)=F_0(y,t).\label{rho0}
\end{align}
 Substituting \eref{qythetat} in \eref{fp-scaled}, and integrating over $\theta$, we find that $F_0(y,t)$ satisfies,
\begin{align}
\frac{\partial F_0(y,t)}{\partial t}=-\frac{\sqrt{2}}{\varepsilon}\frac{\partial F_1}{\partial y},
\label{fmequal0}
\end{align}
 which involves $F_1(y,t)$. To find $F_1(y,t)$, we require $F_2(y,t)$ and so on. In general, substituting \eref{qythetat} in \eref{fp-scaled},
multiplying both sides by $\cos(m\theta)$ and integrating over $\theta$, we get,
\begin{align}
\left(\varepsilon^2\frac{\partial}{\partial t}+1\right)F_m=-\frac{\varepsilon}{\sqrt{2}}\frac{\partial}{\partial y}\left[F_{m-1}(y,t)+F_{m+1}(y,t) \right]~~\text{for }m>0.
\label{fmgreat0}
\end{align}

To extract the long time behavior systematically, we expand $F_m(y,t)$ as an infinite series in the dimensionless parameter $\varepsilon/\sqrt{t}$,
\begin{align}
F_m(y,t)=\sum_{k=0}^{\infty}\varepsilon^k A_m^k(y,t),\label{series}
\end{align}
where the factors $t^{-k/2}$ are absorbed in the series coefficients $A_m^k(y,t)$. Evidently, $A_m^k(y,t)=0$ for $k<0$.
Note that, since \eref{fmequal0} and \eref{fmgreat0} are invariant under the transformation $(y,\varepsilon)\to(-y,-\varepsilon)$, $F_m(y,t)$ must also be invariant under the same transformation, which, in turn, implies that, 
\begin{align}
A_m^k(-y,t)=(-1)^k A_m^k(y,t).\label{parity}
\end{align}

\begin{figure}
\centering \includegraphics[width=0.5\hsize]{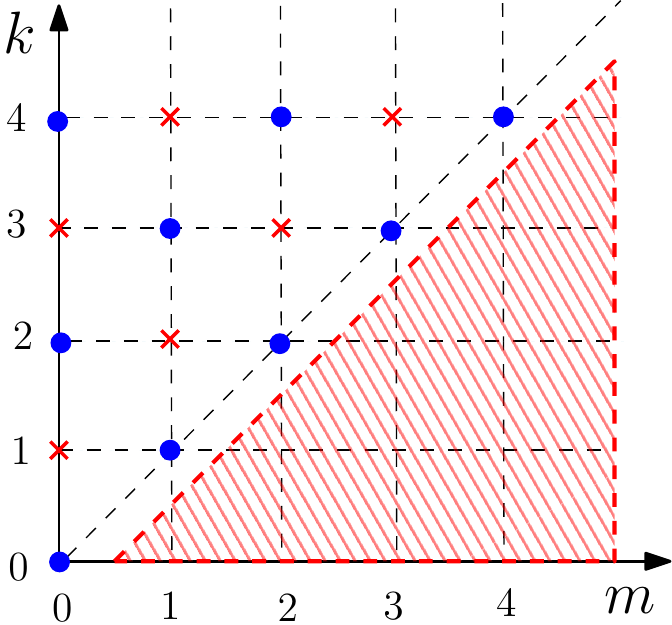}
\caption{Graphical representation of the non-zero $A_m^k(y,t)$ in the $(m,k)$ lattice following \eref{a0k} and \eref{amk}. The filled blue discs denote the lattice points at which $A_m^k(y,t)$ is non-zero, while the red crosses denote the points corresponding to $A_m^k(y,t)=0$. The red striped region covers the lattice points with $m>k$, for which $A_m^k(y,t)$ vanish.}
\label{f:am}
\end{figure}

Substituting this expansion in \eref{fmequal0} and \eref{fmgreat0} and comparing the terms of order $\varepsilon^k$ on both sides, we get,
\begin{align}
\frac{\partial A_0^{k-2}}{\partial t}=-\sqrt{2}\frac{\partial A_1^{k-1}}{\partial y},\label{a0k}
\end{align}
and
\begin{align}
\frac{\partial A_m^{k-2}}{\partial t}=-\frac{1}{\sqrt{2}}\frac{\partial}{\partial y}\left(A_{m-1}^{k-1}+ A_{m+1}^{k-1}  \right)-A_m^k~~\text{for }m>0.\label{amk}
\end{align}
 Evidently, 
 \begin{align}
A_m^0(y,t)=\delta_{m,0}A_0^0.\label{am0delta}
\end{align}
Again, putting $k=1$ in \eref{amk}, and using the above relation we have,
\begin{align}
A_m^1(y,t)=-\frac{\delta_{m,1}}{\sqrt{2}}\frac{\partial A_{0}^0}{\partial y}.\label{am1delta}
\end{align}
Smilarly one can proceed for $k=2,\,3,\dotsc,$ and it follows from the structure of \eref{amk} along with \eref{am0delta} and \eref{am1delta} that $A_m^k=0$ for $k<m$, which is illustrated graphically in \fref{f:am}.

Since the marginal position distribution $F_0(y,t)$ is symmetric in $y$, it follows from \eref{parity} that $A_0^k(y,t)=0$ for odd $k$, and we can write from \eref{series},
\begin{align}
F_0(y,t)=\sum_{k=0}^{\infty}\varepsilon^{2k} A_0^{2k}(y,t).\label{seriesf0}
\end{align}
Following \eref{a0k}, the coefficients $A_0^{2k}(y,t)$ satisfy the differential equation,
\begin{align}
\frac{\partial A_0^{2k}}{\partial t}=-\sqrt{2}\frac{\partial A_1^{2k+1}}{\partial y},\label{a02k}
\end{align}
which, in turn, requires $A_m^{k}(y,t)$ with $m>1$. It should be mentioned here that the conditions $A_0^0(y,t)\neq 0$ and $A_0^1(y,t)=0$ lead to [see \fref{f:am}], 
\begin{align}
A_m^k(y,t)=0 \text{ for all odd }(m+k).
\end{align}


Now, we proceed with the explicit evaluation of $A_0^{2k}(y,t)$ systematically.
The equation for the leading order term $A_0^0(y,t)$ is obtained by putting $k=0$ in \eref{a02k},
\begin{align}
\frac{\partial A_0^{0}}{\partial t}=-\sqrt{2}\frac{\partial A_1^{1}}{\partial y},\label{a001}
\end{align}
where $A_1^1(y,t)$ has to be obtained in terms of $A_0^0(y,t)$ to get a closed form equation. By substituting $k=m$ in \eref{amk}, one can, in fact, find the  general relation, 
\begin{align}
A_m^m(y,t)=-\frac{1}{\sqrt{2}}\frac{\partial A_{m-1}^{m-1}}{\partial y}=\left(-\frac{1}{\sqrt{2}}\right)^m\frac{\partial A_{0}^{0}}{\partial y}.\label{amm}
\end{align}
Thus, using $A_1^1(y,t)=-\frac{1}{\sqrt{2}}\frac{\partial A_{0}^{0}}{\partial y}$ in \eref{a001}, yields a diffusion equation for $A_0^0(y,t)$,
\begin{align}
\frac{\partial A_0^{0}}{\partial t}=\frac{\partial^2 A_0^{0}}{\partial y^2}.
\end{align}
The normalized marginal position distribution to order $\varepsilon^0$ is therefore given by,
\begin{align}
A_0^0(y,t)=\frac{1}{\sqrt{4\pi t}}\exp\left(-\frac{y^2}{4t}\right).
\end{align}
At the next order, putting $k=1$ in \eref{a02k}, we have,
\begin{align}
\frac{\partial A_0^{2}}{\partial t}=-\sqrt{2}\frac{\partial A_1^{3}}{\partial y}.
\end{align}
To get a closed differential equation for $A_0^2(y,t)$, we need to express the right hand side of the above equation in terms of $A_0^2(y,t)$ itself, or the already known function $A_0^0(y,t)$. To this end, we put $m=1$ and $k=3$ in \eref{amk},
\begin{align}
A_1^3=-\frac{\partial A_1^1}{\partial t}-\frac{1}{\sqrt{2}}\left(\frac{\partial A_{0}^{2}}{\partial y}+\frac{\partial A_{2}^{2}}{\partial y}  \right).
\end{align}
The unknown terms on the right hand side of the above equation, namely, $A_1^1(y,t)$ and $A_2^2(y,t)$ can be expressed in terms of $A_0^0(y,t)$ using \eref{amm}. Thus, we have an inhomogeneous diffusion equation,
\begin{align}
\left[\frac{\partial }{\partial t}-\frac{\partial^2 }{\partial y^2}\right]A_0^{2}(y,t)=S_2(y,t),\label{a02}
\end{align}
where the source term $S_2(y,t)$ is given in terms of $A_0^0(y,t)$ as,
\begin{align}
S_2(y,t)=-\frac{\partial^2 }{\partial y^2}\left(\frac{\partial }{\partial t}-\frac 12\frac{\partial^2 }{\partial y^2}\right)A_0^{0}(y,t).\label{s2}
\end{align}
Owing to the diffusive nature of $A_0^0(y,t)$ and the structure of \eref{a02} and \eref{s2}, we anticipate the scaling form,
\begin{align}
A_0^{2}(y,t)=\frac{1}{t}q_{2}\left(\frac{y}{\sqrt{4t}}\right)\frac{e^{-y^2/(4t)}}{\sqrt{4\pi t}}.\label{scalingform02}
\end{align}
Using this scaling form in \eref{a02}, we get an inhomogeneous Hermite differential equation for $q_2(z)$,
\begin{align}
q_2''(z)-2zq_2'(z)+4q_2(z)=s_2(z),\qquad\text{with }s_2(z)=\frac 32-6z^2+2z^4.
\end{align}
The solution of the above differential equation is given by,
\begin{align}
q_2(z)=c_2H_2(z)+\frac{z^2}{4}(3-2z^2),\label{q2_1}
\end{align}
where $H_n(z)$ is the Hermite polynomial of order $n$ and $c_2$ is an arbitrary constant. The normalization condition of the marginal position distribution $\int_{-\infty}^{\infty}dy F_0(y,t)=1$ is satisfied for all values of $c_2$. Therefore, to determine $c_2$ we compare the $O(\varepsilon^2)$ term of the second moment computed from \eref{seriesf0}, with $-2\varepsilon^2$, obtained from \eref{m20t_f}. From \eref{seriesf0} the $O(\varepsilon^2)$ term of the second moment is given by, 
\begin{align}
\varepsilon^2\int_{-\infty}^{\infty}dy\,y^2 A_0^2(y,t)=4\varepsilon^2\int_{-\infty}^{\infty}dz\, z^{2}\,q_{2}(z)\frac{e^{-z^2}}{\sqrt{\pi}}=\varepsilon^2\left(8c_2-\frac 32\right),
\end{align}
where we have used \eref{scalingform02} and \eref{q2_1} successively. Comparing the right hand side with $-2\varepsilon^2$, we get $c_2=-1/16$, which, in turn, leads to,
\begin{align}
q_2(z)=\frac18(1+4z^2-4z^4).
\end{align}
Similarly, we determine the next higher order subleading corrections $A_0^4(y,t)$, $A_0^6(y,t)$ and so on. They satisfy the inhomogeneous diffusion equation,
\begin{align}
\left[\frac{\partial }{\partial t}-\frac{\partial^2 }{\partial y^2}\right]A_0^{2k}(y,t)=S_{2k}(y,t),\label{a02k_diff}
\end{align}
where the source term $S_{2k}(y,t)$ depends on the previous order solutions. For example,
\begin{align}
S_4(y,t)&=-\frac{\partial^2 }{\partial y^2}\left(\frac{\partial}{\partial t}-\frac 12\frac{\partial^2}{\partial y^2}\right)A_0^2(y,t)+\frac{\partial^2}{\partial y^2}
\left(\frac{\partial}{\partial t^2}-\frac 32\frac{\partial^3}{\partial y^2 \partial t}+\frac 12\frac{\partial^4}{\partial y^4}\right)A_0^0(y,t),\label{S2z}\\
S_6(y,t)&=\frac{\partial^2 }{\partial y^2}\left(\frac{\partial }{\partial t}-\frac 12 \frac{\partial^2 }{\partial y^2}   \right)A_0^4(y,t)+\frac{\partial^2 }{\partial y^2}\left(\frac{\partial^2 }{\partial t^2}-\frac 32\frac{\partial^3 }{\partial y^2\partial t} +\frac 12 \frac{\partial^4 }{\partial y^4} \right)A_0^2(y,t)\cr
&-\frac{\partial^2 }{\partial y^2}\left(\frac{\partial^3 }{\partial t^3}+3\frac{\partial^4 }{\partial y^2 \partial t^2} -\frac 52 \frac{\partial^5 }{\partial y^4 \partial t}+\frac{5}{16}\frac{\partial^6}{\partial y^6} \right)A_0^0(y,t).\label{S4z}
\end{align}
Using the scaling forms,
\begin{align}
A_0^{2k}(y,t)=&\frac{1}{t^k}q_{2k}\left(\frac{y}{\sqrt{4t}}\right)\frac{e^{-y^2/(4t)}}{\sqrt{4\pi t}},\label{scalingform}\\
\intertext{and} S_{2k}(y,t)=&\frac{1}{t^{k+1}}s_{2k}\left(\frac{y}{\sqrt{4t}}\right)\frac{e^{-y^2/(4t)}}{\sqrt{4\pi t}},\label{scalingforms2}
\end{align}
in \eref{a02k_diff} we get an inhomogeneous Hermite differential equation at each order as,
\begin{align}
q_{2k}''(z)-2zq_{2k}'(z)+4k\,q_{2k}(z)=s_{2k}(z).\label{hermiteq2k}
\end{align}

In general, the physically admissible solution to \eref{hermiteq2k} is given by,
\begin{align}
\fl \qquad~~\qquad q_{2k}(z)=c_{2k}\, U_{2k}(z) +(-1)^k\frac{k!}{(2k)!}\int_0^z dy\,e^{-y^2}\,\Big[ V_{2k}(z)\,U_{2k}(y)-U_{2k}(z)\,V_{2k}(y)\Big]s_{2k}(y),\label{gensolnq2k}
\end{align}
where $U_{2k}(z)=H_{2k}(z)$ is the Hermite polynomial of order $2k$ and $V_{2k}(z)=z\, {}_1F_1\left(1/2-k,3/2,z^2\right)$ is the confluent hypergeometric function. These are the two independent solution of the homogeneous Hermite differential equation $q_{2k}''(z)-2zq_{2k}'(z)+4k\,q_{2k}(z)=0$.
The arbitrary constant $c_{2k}$ in \eref{gensolnq2k} is determined by comparing the coefficient of $\varepsilon^{2k}$ of $M(2k,0,t)$ obtained in the previous section with
\begin{align}
4^k\int_{-\infty}^{\infty}dz\, z^{2k}q_{2k}(z)\frac{e^{-z^2}}{\sqrt{\pi}},
\end{align}
obtained from \eref{seriesf0}, using \eref{scalingform} and \eref{gensolnq2k}. 
We next explicitly compute the corrections for $k=2,\,3$. 

For $k=2$, we have,
\begin{align}
s_4(z)=&\frac{15}{16}+\frac{15 z^2}2-\frac{45 z^4}2+10 z^6-z^8,
\end{align}
which leads to,
\begin{align}
q_4(z)=\frac{1}{128}\bigl(9+24 z^2+72 z^4-96 z^6+16 z^8\bigr).
\end{align}
Using $q_2(z)$ and $q_4(z)$, for $k=3$, we obtain,
\begin{align}
s_6(z)=&\frac{1}{256}\bigl(315 + 1260 z^2 + 6300 z^4 - 16800 z^6 + 8400 z^8 - 1344 z^{10} + 64 z^{12}\bigr),
\end{align}
which, after determining $c_6$, gives,
\begin{align}
q_6(z)=\frac{1}{3072}\bigl(225 + 540 z^2 + 900 z^4 + 2400 z^6 - 3600 z^8 + 960 z^{10} - 64 z^{12}\bigr).
\end{align}
Proceeding similarly, one can systematically calculate the higher order corrections. Finally, remembering that the isotropy of two-dimensional position distribution, the radial distribution of the RTP in the diffusive scaling limit can be expressed in the universal form~\cite{Santra_2022},
\begin{align}
P_r(r,t)=\frac{1}{\sqrt{\pi D_\rt t}}\exp\left(-\frac{r^2}{4D_\rt t} \right)\sum_{k=0}^{\infty}\left(\gamma t\right)^{-k}\, q_{2k}\left(\frac{r}{\sqrt{4D_\rt t}}\right), \label{exactseries}
\end{align}
with $\int_0^\infty\, P_r(r,t)\,dr=1$.
Note that, for a passive Brownian paricle $q_{2k}(z)=0$ for $k> 0$. Therefore, the emergence of the non-trivial polynomials $\{q_{2k}(z);k> 0\}$ is solely due to the active nature of the underlying dynamics. Moreover, the form of $q_{2k}(z)$ depends on the specific active dynamics~\cite{Santra_2022}. These signatures of activity in the diffusive scale [i.e., involving typical trajectories showing fluctuations $O(\sqrt{t})$] are easier to observe in experiments, in comparison to the large deviation form which encodes rare fluctuations of $ O(t)$.

\section{Effect of translational diffusion}\label{sec:thermal}
We have ignored the effect of thermal fluctuations in our calculations so far. In this section, we investigate the effect of a thermal translational noise $\vec{\eta}=(\eta_1,\eta_2)$. In that case, the Langevin equation \eref{e:langevin} changes to,
\begin{align}
\dot{\vec x}=\vec{v}(t)+\sqrt{2D}\,\vec{\eta}(t),\label{e:langevin2}
\end{align}
where $\la\eta_i\ra=0$, $\la\eta_i(t)\eta_j(t')\ra=\delta_{ij}\delta(t-t')$ and $D$ denotes the translational diffusion coefficient.

Since $\vec{v}(t)$ and $\vec{\eta}(t)$ are two independent random noises, the distribution of the position $\vec{x}(t)=(x_1(t),x_2(t))$ can be expressed as a convolution,
\begin{align}
P_D(\vec{x},t)=\int d\vec{x'}P_a(\vec{x'},t)P_p(\vec{x}-\vec{x'},t),
\end{align}
where $P_a(\vec{x},t)$ denotes the distribution of the process $\vec{x}(t)=\int_0^t \vec{v}(s)\,ds$ and $P_p(\vec{x'},t)$ is the distribution of the diffusion process $\vec{x'}(t)=\int_0^t \vec{\eta}(s)\,ds$. In particular, for the marginal distribution of $x_1$, we have,
\begin{align}
P_D(x_1,t)=\int dx'P_a(x',t)P_p(x_1-x',t),\label{convo1}
\end{align}
where $P_p(x_1,t)$ satisfies the diffusion equation,
\begin{align}
\frac{\partial}{\partial t}P_p(x_1,t)=D\frac{\partial^2 }{\partial x_1^2}P_p(x_1,t).\label{eq:pp}
\end{align} 
It follows from the analysis in the previous sections (see \eref{rho0}, \eref{seriesf0} and \eref{a02k_diff} for example), that,
\begin{align}
P_a(x_1,t)=\sum_{n=0}^{\infty}\gamma^{-n}p_n(x_1,t)\label{eq:paexpand},
\end{align}
where $p_n(x,t)$ satisfies the inhomogeneous diffusion equation,
\begin{align}
\left[\frac{\partial }{\partial t}-D_\rt\frac{\partial^2 }{\partial x^2}\right]p_n(x,t)=\mathcal{S}_n(x,t),\label{eq:pn_diff}
\end{align}
where $\mathcal{S}_0(x,t)=0$ and $\mathcal{S}_n(x,t)$ for $n>0$ is related to $S_{2n}(y,t)$ [appearing in \eref{a02k_diff} for the scaled variable $x/\sqrt{D_\rt}$].
%
%
%
%

Using \eref{eq:paexpand} in \eref{convo1}, we find that the marginal position distribution in the presence of translational noise is given by, 
\begin{align}
P_D(x_1,t)=\sum_{n=0}^{\infty}\gamma^{-n}P_D^{(n)}(x_1,t),\label{rhod}
\end{align}
where,
\begin{align}
P_D^{(n)}(x_1,t)=\int_{-\infty}^{\infty}dx'\,p_n(x',t)\,P_p(x_1-x',t).\label{rhod2}
\end{align}
To study the time evolution of the probability distribution $P_D^{(n)}(x_1,t)$, we take a derivative of \eref{rhod2} with respect to time to obtain,
\begin{align}
\frac{\partial P_D^{(n)}(x_1,t)}{\partial t}&=\int_{-\infty}^{\infty}dx'\,\Big[\frac{\partial p_n(x',t)}{\partial t}\,P_p(x_1-x',t)
+p_n(x',t)\,\frac{\partial P_p(x_1-x',t)}{\partial t}\Big].
\end{align}
Now, we first use \eref{eq:pp} and \eref{eq:pn_diff} in the above equation to replace the time derivative in terms of the spatial derivatives. Subsequently, integrating by parts and using $\partial_{x_1} P_p(x_1-x')=-\partial_{x'}P_p(x_1-x')$, we find that $P_D^{(n)}(x_1,t)$ satisfies the inhomogeneous diffusion equation,
\begin{align}
\left[\frac{\partial }{\partial t}-\Lambda\frac{\partial }{\partial x_1^2}\right]P_D^{(n)}(x_1,t)=\tilde{\mathcal{S}}_n(x_1,t),
\end{align}
where $\Lambda=(D+D_\rt)$ and
\begin{align}
\tilde{\mathcal{S}}_n(x_1,t)=\int_{-\infty}^{\infty}dx'\, \mathcal{S}_{n}(x',t)P_p(x_1-x',t).
\end{align}
Thus, the translational diffusion modifies the effective diffusion coefficient, as well as the source functions $\{\tilde{\mathcal{S}}_n(x_1,t)\}$. 
\begin{figure}
\centering\includegraphics[width=0.65\hsize]{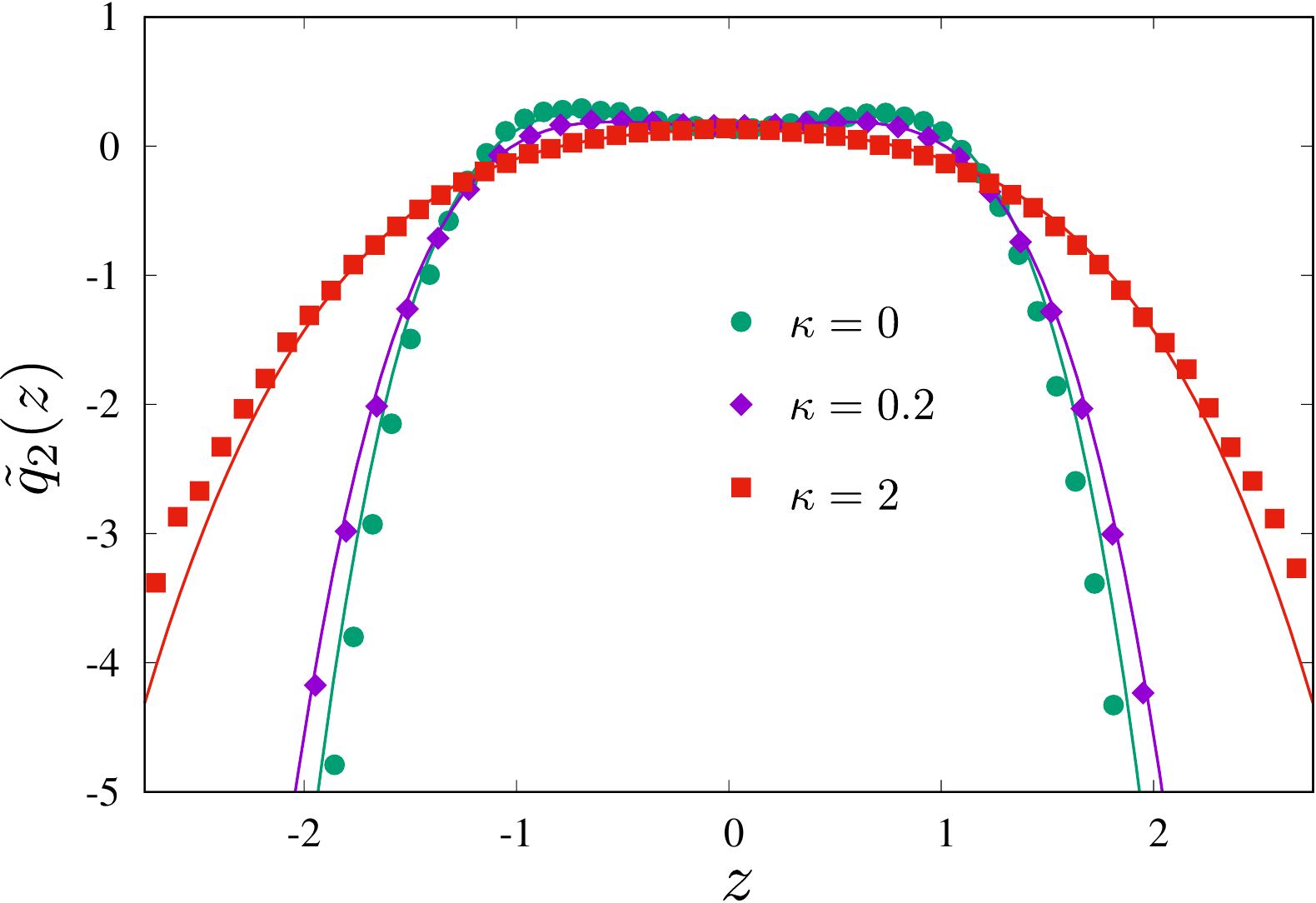}
\caption{Plot of $\tilde{q}_{2k}(z)$ for different values of the translational diffusion coefficient $D$ (with $\gamma=1$ and $v_0=1$), showing the shape transition at the origin from a local minimum for smaller values of $\kappa$ to a maximum for large $\kappa$. The case $\kappa=0$ corresponds to the case of no translational diffusion. The symbols denote $\tilde{q}_{2k}(z)$ extracted from numerical simulations with $t=10$, while the solid lines correspond to \eref{tildeq2}. }\label{f:q2}.
\end{figure}
Therefore, skipping details, \eref{rhod} becomes,
\begin{align}
P_D(x_1,t)=\frac{1}{\sqrt{4\pi \Lambda t}}\exp\left(-\frac{x_1^2}{4\Lambda t} \right)\sum_{k=0}^{\infty}\left(\gamma t\right)^{-k}\,\tilde{q}_{2k}\left(\frac{x_1}{\sqrt{4\Lambda t}}\right),\label{pdiff}
\end{align}
where $\tilde{q}_{0}(z)=1$, and $\tilde{q}_{2k}(z)$ for $k>0$ quantify the corrections to the leading order Gaussian distribution. The polynomials  $\tilde{q}_{2k}(z)$ can be obtained by comparing the coefficients of $(\gamma t)^{-k}$ in the above equation and \eref{convo1}. In general, the correction polynomials $\tilde{q}_{2k}(z)$ can be written in terms of the corresponding correction polynomials $q_{2k}(z)$ [see \eref{exactseries}] as,
\begin{align}
\tilde{q}_{2k}(z)=\int_{-\infty}^{\infty}dw\,\frac{\exp\left(-w^2/\kappa\right)}{\sqrt{\pi \kappa}}\,q_{2k}\left(\frac{z-w}{\sqrt{1+\kappa}}\right),\label{tildeq2k}
\end{align}
where $\kappa=D/D_\rt$ denotes the ratio of the translational diffusion and the effective diffusion coefficient of the RTP in the absence of the translational diffusion. Equation \eref{tildeq2k} is a very general relation, that relates the correction polynomials in the presence of translational diffusion to the ones in absence of translational diffusion for any active particle model.
The integral in \eref{tildeq2k} can be evaluated exactly for any polynomial function $q_{2k}(z)$. For example the first few terms are given by,
\begin{align}
\tilde{q}_2(z)=&\frac{1}{8(1+\kappa)^2}\left(1+4\kappa+4(1-2\kappa)z^2-4z^4\right),\label{tildeq2}\\
\tilde{q}_4(z)=&\frac{1}{128(1+\kappa)^4}
\left[ 9+48\kappa+144\kappa^2+24z^2(1+12\kappa-24\kappa^2)
\right.\cr
&\left.\qquad\qquad+
24z^4(3-24\kappa+8\kappa^2)-32z^6(3-4\kappa)+16z^8\right],\\
\tilde{q}_6(z)=&\frac{1}{3072(1+\kappa)^6} \left(5+36 \kappa+120 \kappa ^2+320 \kappa ^3 +540z^2(1+10\kappa+80\kappa^2-160\kappa^3)\right.\cr &\left.\qquad+900z^4(1+24\kappa-144\kappa^2+64\kappa^3)+480z^6(5-90
\kappa+120\kappa^2-16\kappa^3)\right.\cr
&\left.
\qquad\qquad-720z^8(5-20\kappa+8\kappa^2)+192z^{10}(5-6\kappa)-64z^{12}\right).
\end{align}
\Fref{f:q2} shows the leading order correction polynomial $\tilde{q}_2(z)$ in the absence and presence of translational diffusion. It is interesting to note that in the presence of translational diffusion $\tilde{q}_2(z)$ undergoes a shape transition --- $z=0$ is a local minimum of $\tilde{q}_2(z)$ for $\kappa<1/2$, whereas it becomes a maximum for $\kappa> 1/2$. We recall that, in the absence of translational diffusion ($\kappa=0$) $q_2(z)$ always exhibits a minimum at $z=0$. The higher order corrections $\tilde{q}_4(z)$, $\tilde{q}_6(z)$, etc. also undergo similar shape transitions at the origin, albeit at progressively higher values of  $\kappa$.


\section{Conclusion}\label{conclusion}
We use the perturbative procedure developed in~\cite{Santra_2022} to calculate the long-time position distribution of a run-and-tumble particle in two dimensions with propulsion speed $v_0$, tumbling rate $\gamma$, and translational diffusion coefficient $D$. For simplicity, we consider the initial orientation to be isotropic, for which the position distribution also remains isotropic at all times. To understand the long-time behavior of this isotropic position distribution, starting from the Fokker-Planck equation, we show that the long-time marginal position distribution admits a series solution in powers of the dimensionless parameter $(\gamma t)^{-1}$. We find that the leading order contribution to the position distribution satisfies a diffusion equation with an effective diffusion constant $\Lambda=D+D_\rt$, where $D_\rt=v_0^2/(2\gamma)$ is the effective diiffusion coefficient in the absence of translational diffusion ($D=0$).
The subleading corrections satisfy an inhomogeneous diffusion equation where the inhomogeneous term, at each order, depends on the previous order solutions. In particular, the distribution of the scaled radial distance $R=r/\sqrt{4\Lambda t}$ can be expressed as,
$p_D(R,t)=\pi^{-1/2}e^{-R^2}\sum_{k=0}^{\infty}\left(\gamma t\right)^{-k}\, \tilde{q}_{2k}\left(R\right)$, where $\tilde{q}_{2k}(R)$ is a polynomial of order $4k$ that depends on the dimensionless parameter $\kappa=D/D_\rt$. It turns out that $\tilde{q}_{2k}(R)$ can be expressed in terms of $q_{2k}(R)$, the corrections in the absence of translational diffusion ($\kappa=0$), which satisfies inhomogeneous Hermite differential equations at each order. We illustrate the procedure by explicitly calculating the first few corrections $q_{2k}(R)$ and $\tilde{q}_{2k}(R)$. 
As a part of this procedure, we develop a recursive formalism for computing the correlation functions $\la y^k\cos(n\theta)\ra$ exactly in the absence of translational diffusion. In particular, we obtain a closed-form expression for $\la y^k\cos(k\theta)\ra$.
 While the leading order universal Gaussian behavior of the position distribution is expected from the central limit theorem, our work brings out the universal nature of the subleading corrections to the Gaussian, as proposed in our previous work~\cite{Santra_2022}. 
 
An obvious question is if a similar perturbative technique can be used to study some other important physical observables, like first-passage time~\cite{Mori2020_prl} and convex-hull~\cite{hartmann2020convex,singh2022mean}, of two-dimensional RTPs.
 It would also be interesting to see whether the predicted universal properties still hold if the underlying orientation dynamics is nonequilibrium, for example, chiral active motion. Another relevant question is whether the universal structure survives if the time between the consecutive tumble events are taken from more generalized waiting-time distributions, but still having a characteristic time-scale. 

\ack U. B. acknowledges support from the Science and Engineering Research Board (SERB), India, under a Ramanujan Fellowship (Grant No. SB/S2/RJN-077/2018).

\appendix

\section{Solution of the recursion relation for the correlation function $M(k,k,t)$}\label{sec:mkkt}

In this appendix, we provide the derivation of the correlation function $M(k,k,t)$ given in \eref{mkkt} in the main text. Equation \eref{mkkt:1} is of the form,
\begin{align}
f(k,t)=\frac{k}{\sqrt{2/b}}\int_0^t ds\, e^{-b(t-s)}\,f(k-1,s),\label{app:fkt}
\end{align}
where $f(k,t)=M(k,k,t)$ and $b=1/\varepsilon^2$. To solve the above integral recursion relation, we take a Laplace transform with respect to $t$ on both sides of \eref{app:fkt} to get,
\begin{align}
\tilde{f}(k,\lambda)=\int_0^{\infty}dt\, e^{-\lambda t}\,f(k,t)&=\frac{k}{\sqrt{2/b}}\int_0^{\infty}dt\, e^{-\lambda t}\int_0^t ds e^{-b(t-s)}\,f(k-1,s),\cr
&=\frac{k}{\sqrt{2/b}}\,\frac{1}{\lambda+b}\,\tilde{f}(k-1,\lambda).
\end{align}
The above equation is a simple algebraic recursion relation with the initial condition $\tilde{f}(0,\lambda)=1/\lambda$ [since $f(0,t)=M(0,0,t)=1$], and can be solved to get,
\begin{align}
\tilde{f}(k,\lambda)&=\frac{k!}{(2/b)^{k/2}\,\lambda\,(\lambda+b)^k}.
\end{align}
The generating function can be inverted exactly to yield,
\begin{align}
f(k,t)&=\frac{k!}{(2b)^{k/2}}\Big(1-e^{-bt}\sum_{n=0}^{k-1}\frac{b^n}{n!}\Big).
\end{align}
Subtituting $b=1/\varepsilon^2$, we get \eref{mkkt}, quoted in the main text.

\section{Extraction of $q_{2k}(z)$ from the exact solution}\label{app:exact}
The Fokker-Planck equation \eref{pxthetat} has been exactly solved in \cite{Santra2020} to obtain the marginal position distribution as,
\begin{align}
P(x_1,t)=\frac{\gamma\, e^{-\gamma t}}{2v_0}\left[L_0\left(\frac{\gamma}{v_0}\sqrt{v_0^2t^2-x_1^2}\right)+I_0\left(\frac{\gamma}{v_0}\sqrt{v_0^2t^2-x_1^2}\right)\right]+\frac{e^{-\gamma t}}{\pi \sqrt{v_0^2t^2-x_1^2}}  ,\label{exactsol}
\end{align}
where $L_0(u)$ and $I_0(u)$ denote the modified Struve function and modified Bessel function of the first kind, respectively. The last term, which decays exponentially in time, characterizes the ballistic spread and can be ignored in the long-time diffusive regime. In this regime, substituting $v_0=\sqrt{2\gamma D_\rt}$, we get the distribution of the scaled variable $z=x_1/\sqrt{4D_\rt t}$ as,
\begin{align}
P(z,t)=\sqrt{\frac{\gamma t}{2}}\, e^{-\gamma t}\left[L_0\left(\gamma t\sqrt{1-\frac{2z^2}{\gamma t}}\right)+I_0\left(\gamma t\sqrt{1-\frac{2z^2}{\gamma t}}\right)\right].\label{scaledexact}
\end{align}
For large $\gamma t$, the arguments of $L_0$ and $I_0$ are large and
\begin{align}
L_0(u)=I_0(u)=&\frac{e^u}{\pi\sqrt{2u}}\int_0^u \frac{d\tau}{\sqrt{\tau}}e^{-\tau}\left(1-\frac{\tau}{ 2u}\right)^{-1/2}\text{~~as~~}u\to\infty\cr
=&\frac{e^u}{\pi\sqrt{2u}}\sum_{n=0}^{\infty}\frac{u^{-n}}{n!}\left[\frac{(2n-1)!!}{2^n}\right]^2,\label{l0i0_1}
\end{align}
where we have used the expansion of $(1-z)^{-1/2}=(2n-1)!!\,z^n/(2^n n!)$ and taken the upper limit of integration to $\infty$.
The first few terms are given by,
\begin{align}
L_0(u)=I_0(u)=\frac{e^u}{\sqrt{2\pi}}\left[\frac{1}{u^{1/2}}+\frac{1}{8u^{3/2}}+\frac{9}{128u^{5/2}}+\frac{75}{1024u^{7/2}}+O\left(\frac{1}{u^{9/2}}\right]\right).\label{l0io}
\end{align}
Using \eref{l0io} in \eref{scaledexact}, the corrections $q_{2k}(z)$ can be obtained and match exactly with the ones obtained in Sec.~\ref{secpdf}.

\section*{References}
\bibliography{ref}

\end{document}